\documentclass[epj]{svjour}
\begin{document}

\newcommand{\hs}{\hspace{1cm}}
\newcommand{\phib}{\bar{\phi}}
\newcommand{\phiflat}{\phi_{flat}}
\newcommand{\si}{\sigma}
\newcommand{\LA}{\Lambda}
\newcommand{\DE}{\Delta}
\newcommand{\la}{\lambda}
\newcommand{\al}{\alpha}
\newcommand{\be}{\beta}
\newcommand{\ga}{\gamma}
\newcommand{\xs}{x^*}

\title{Inwardly Curved Polymer Brushes : Concave is not Like Convex}

\author{M. Manghi\inst{1} \and M. Aubouy\inst{1} \and
C. Gay\inst{2} \and C. Ligoure\inst{2,3}}

\institute{Groupe Th\'eorie, SI3M, UMR 5819 (CEA-CNRS-Univ. J.
Fourier), D\'{e}partement de Recherche Fondamentale sur la
Mati\`{e}re Condens\'{e}e, CEA-Grenoble, 38054 Grenoble cedex 9,
France, \email{mmanghi@cea.fr, maubouy@cea.fr} \and Laboratoire
CNRS - Elf Atochem (UMR 167), B.P.108, 92303 Levallois-Perret
cedex, France, \email{cgay@pobox.com} \and GDPC CC 26,
Universit\'e' Montpellier 2, 34095 Montpellier Cedex 05, France,
\email{ligoure@gdpc.univ-montp2.fr}}

\date{\today}

\abstract{ Inwardly curved polymer brushes are present in
cylindrical and spherical micelles or in membranes tubes and
vesicles decorated with anchored polymers, and influence their
stability. We consider such polymer brushes in good solvent and
show that previous works, based on a self-similar concentric
structure of the brush, do not describe the most stable structure.
We use scaling laws to derive very simply the leading term of the
free energy in the high curvature limit, where the osmotic
pressure is the relevant physical ingredient. We also derive the
complete conformation at all curvatures using a self-consistent
field approach. The free energy is computed therefrom using a
local scaling description}


\maketitle


\section{Introduction.}


Various types of polymer layers have been studied in recent years.
The most common situations include polymer brushes (where polymers
molecules are end-tethered to a repulsive surface or interface),
reversible adsorption from a dilute polymer solution, irreversible
adsorption from a melt. These are important in the field of
adhesion, wetting, lubrication, colloid stabilization
\cite{napper,holbook,PGG87ASV}. Different theories have been
proposed to describe the structure and the properties of the
polymer coatings as a function of the mode of attachment, the
quality of the solvent, etc. However, concerning the geometry of
the surface which supports the layer, most of the work is
concerned with flat or convex interfaces
\cite{McCONNELL,HALPERIN}. This is in contrast with a number of
situations where polymers chains are attached to concave surfaces.
For example, diblock copolymers can form spherical micelles in a
selective solvent. The core of these is made of the segregated
moeties and can be pictured as a dense layer of chains ``grafted''
to the inner surface of a sphere. Recently, nine different phases
have been found in a ternary system of poly(ethylene
oxide)/poly(propylene oxide) amphiphilic block copolymer, water
and oil (selective solvents). Six of these phases are made of
cylindrical and spherical normal micelles (water-in-oil) or
reverse micelles (oil-in-water)~\cite{Alexandridis}. Lattice Monte
Carlo simulations were performed to study chain conformations in
spherical cavities~\cite{Prochazka}. Polymers attached to fluid
membranes provide another interesting situation. Indeed, in
biological systems, lipid bilayers are often decorated by a large
number of macromolecules. Inspired by biological structures,
simplified model systems of polymers attached to lipid bilayers
have been studied experimentally
\cite{DECHER,SIMON,BLUME,STAVANS}. It was shown that anchoring
polymers induces dramatic changes in the vesicles \cite{DECHER} or
multilamellar membrane tubes \cite{STAVANS} shapes. In general,
the polymers will be attached to both sides, i.e. to both
monolayers of the lipid or surfactant bilayer. Anchored polymers
also stabilize the equilibrium formation of small unilamellar
surfactant vesicles \cite{JOANNIC} over a wide range of
composition and temperature range: the vesicles are spontaneously
spherical and monodisperse and their radius is small (comparable
to the natural size of the anchored polymers). This spontaneous
tendency for bilayers to bend so to form vesicles has been
explained by a coupling \cite{SAFRAN} between the curvature and
the asymmetry of grafting polymers on each monolayer \cite{PORTE}.

Polymer adsorption finds extensive applications in modifying
physical and chemical properties of surfaces. In this context,
experiments have highlighted the effect of surface roughness at
the size scale of the grafting polymers on the total amount of
polymer adsorbed \cite{SINGH}. A significant enhancement of
adsorption was observed when the bead to polymer radius is of
order of unity, indicating that polymer brushes will prefer to
adsorb on spherical or cylindrical bumps (i.e. defects of large
{\it positive} curvature), since the deformation energy of the
brush will be lowered in these geometries with respect to the flat
case \cite{LIGOURE}. On the contrary, valleys, (i.e. defects with
large {\it negative} curvature) substantially reduce the grafting.
Such a reduction was understood in terms of topologically reduced
steric hindrance to adsorption. Monte-Carlo computer simulations
were also performed \cite{BALAZS} to study the adsorption of
triblock copolymers on rough surfaces and confirm that grafting on
defects of negative curvature (concave surfaces) is not favoured
because of conformational energy penalty of the anchored polymers.

Surprisingly, even if it was anticipated that in the high
curvature limit, concave and convex brushes should not be
described in a similar way~\cite{LIGOURE}, very few theoretical
works have been devoted to the specific case of inwardly curved
brushes up to date. To our knowledge, only a few theoretical papers
have addressed this problem, two of them in the context of the study of
the elastic properties of polymer-decorated membranes
\cite{Lipowsky,Hristova}, one in the context of shear swelling of
polymer brushes grafted onto convex and concave surfaces
\cite{SEVICK}, two in the context of microemulsion
stabilization by diblock copolymers~\cite{Dan,Govorun}.
E.~Zhulina~\cite{Zhulinapriv}
also pointed out references \cite{Zhulina1,Zhulina2}.
However, most of these authors described the chain conformation and the
free energy of the brush on the concave side of the interface by
using an inverted version of the classical Daoud-Cotton
scaling approach \cite{DAOUDCOTTON}.
In this approach, the structure of the brush is self-similar and
scales linearly with the distance from the center of curvature
(Figure~1). It correctly describes brushes anchored on the {\it
convex} side of interfaces but not on {\it concave} side, because,
as we will show below, this structure is not the most stable one.

The main concern of the present work is to provide a correct
description of the concave brush structure. In order to clarify
the situation of interest, let us briefly review the different
conformation regimes (Figure~2), depending on the number $f$ of
polymer chains (assumed to be linear, flexible, made of identical,
neutral monomers of size $a$), on the number $N$ of monomers per
chain and on the radius $R$ of the inner surface of the sphere.
The number of chains per unit surface area of chains (grafting
density) can be written adimensionally as $\sigma =a^2f/(4\pi
R^2)$. For simplicity reasons, we only treat the case of an
athermal solvent~: the excluded volume parameter is $v\cong 1$ (in
units of $a^3$). We assume that the chains are repelled by the
interface. An important practical issue requiring attention is the
case of flexible membranes with grafted polymers. From this point
of view, we will now express all quantities in terms of the
grafting density, $\sigma$, rather than in terms of the number of
chains, $f$. It allows also an easier comparison with the limit
case of a flat brush. \\

At large sphere radius $R$, if chains are grafted too far apart to
interact, they form separate mushrooms (regime 1 of Figure~2)
whose Flory radius of gyration is $R_F=aN^{3/5}$. They start to
interact when $\sigma \cong (a/R_F)^2$, {\it i.e.}, when~:
\begin{equation}
\sigma \simeq N^{-6/5}
\label{L12}
\end{equation}
For larger values of the grafting density $\sigma $, chains are
stretched in good solvent and form a brush whose height is
essentially the same as in the planar geometry~:
\begin{equation}
L_{flat}\cong aN\sigma^{1/3} \label{extenplan}
\end{equation}
This is regime~2 of Figure~2. It is limited by two phenomena.
First, the grafting density cannot exceed one chain per squared
monomer size, for obvious steric reasons. This condition reads~:
\begin{equation}
\sigma \leq 1
\label{L2NP}
\end{equation}
Secondly, the sphere curvature becomes important when $R\simeq
L_{flat}$, {\it i.e.}, when~:
\begin{equation}
R\simeq aN\sigma^{1/3}
\label{L23}
\end{equation}
At smaller radii, the description of the grafted chain
conformation is not straightforward since both the curvature of
the sphere and the interaction between neighbouring chains are
important. These are regime~3 (where chains are still stretched)
and regime~4 of Figure~2 (where chains are confined). The
separation of these two regimes occurs when the radius is equal to
the natural size of a chain in a semi-dilute solution (see
Section~3.2), {\it i.e.} when
\begin{equation}
R\simeq a\sigma^{-1/7}N^{3/7} \label{L34}
\end{equation}
They are limited by two conditions: one occurs when no more
solvent is present~:
\begin{equation}
R\geq aN\sigma \label{L34NP}
\end{equation}
and the second one when there is only one chain in the sphere~:
\begin{equation}
R\geq a\sigma^{-1/2}
\end{equation}
In regimes~2 and~3, chains are stretched and their conformations
should play an important role in the structure and the energy of
brushes in these regimes. We thus briefly review two usual types of
descriptions.
In scaling descriptions, all chain conformations are identical and
in particular, chain ends are all located at the same distance
from the grafting interface. Conversely, self-consistent field
(SCF) descriptions allow for a spatial distribution of chain ends.
Correlatively, all chain conformations are not identical : at any
given distance from the wall, chains are stretched differently
according to how far from the grafting surface their free ends are
located.

The common feature of both approaches is that the brush
conformation results from the interplay between the two-body
interactions between monomers pertaining to the brush chains
(which tend to stretch the brush further out) and the elastic
stretching of the brush chains (which limits the brush extension).
As long as these two components are well taken into account, both
types of approaches yield essentially similar conclusions.
\newline
Thus, the Daoud-Cotton description of convex brushes, although
based on scaling arguments, yields correct results. Let us
emphasize a few features of such a system. In reality, free chain
ends can sustain no stretching, whereas the grafted ends are under
tension. In the model by Daoud and Cotton, the typical blob size
(see Figure~1a) is larger at greater distances from the wall (it
is proportional to the distance from the center of curvature).
Thus local stretching is weaker far from the wall. This feature
ensures local equilibrium of the chain strands: the gradient in
concentration and thus in two-body interactions tends to push them
away from the wall, whereas the gradient in degree of stretching
tends to pull them back towards the wall. If we focus on the last
blob, the elastic force, given by the Pincus law~\cite{PINCUS},
is $k_BT/\xi(L)$,
where $\xi(L)$ is the size of the last blob. Since it is large,
the elastic force is small and is balanced by the small
concentration gradient at the edge of the brush. The success of
this scaling vision is thus due to the fact that chain-ends are
segregated in the last blob and they sustain almost no stretching.
Hence, the basic ingredients are thus explicitely present in this
scaling description.
\newline
Convexity of the interface is a key feature in this success of the
Daoud-Cotton model~: the planar geometry is a marginal case for
the Alexander-de~Gennes scaling model. It yields essentially the
same results (brush extension and free energy) as the pionneering
SCF description by Semenov~\cite{SEM85} which was further
developed independently by two groups \cite{MWC,ZHULINA}. However,
the monomer concentration profile found in the SCF theory is
parabolic and vanishes at the end of the brush. This is different
from the step profile of the scaling description which imposes a
sharp density gradient at the edge. We might argue that this
gradient balances the elastic restoring force exerted on the
stretched chain and that it ensures the global equilibrium of
chains. But it has been shown that this is not the most stable
structure since the SCF theory yields a lower minimum free energy.

Things get worse in the case of a concave geometry. Two scaling
descriptions can be found in literature~: one, adapted from the
model by Daoud and Cotton~\cite{Lipowsky,Hristova,Dan,Zhulina1},
the other, given by Sevick~\cite{SEVICK}, is based on an
Alexander-de~Gennes brush treated locally to account for a
cylindrical geometry. The essential feature of the ``inverse''
Daoud-Cotton approach is that the blob size is still taken as
proportional to the distance from the center of curvature~:
because the interface is concave, the blob size decreases as we go
further away from the wall (Figure~1b). The essential results of
this model are as follows (Throughout this article, free energies
$F$ are in units of thermal energy $k_BT$)~:

\begin{itemize}

    \item{the thickness of the brush is $L_{DC}\simeq
R[1-(1-\frac{aN\sigma^{1/3}}{R})^{3/5}]$}

    \item{the free energy per chain is $F_{DC}\simeq
-N\sigma^{5/6}\frac{\ln[1-aN\sigma^{1/3}/R]}{aN\sigma^{1/3}/R}$}

    \item{the volume fraction profile is $\phi(r)\simeq
\frac{\sigma^{2/3}}{(1-r/R)^{4/3}}$ where $r$ is the radial
distance to the spherical surface (see Figure 1bis)}

\end{itemize}

Among these predictions, the last one is certainly the most
surprising. It predicts that the volume fraction increases when we
move off from the surface, and notably, at the edge of the brush,
we find $\phi(L)\simeq
\frac{\sigma^{2/3}}{(1-aN\sigma^{1/3}/R)^{4/5}}$. Thus forces
exerted on the last blob (osmotic gradient and elastic force) are
very important. Eventually, the volume fraction diverges at the
center for $R\simeq aN\sigma^{1/3}$. Hence, even if the local
stability inside the brush is fulfilled, as in the convex case, we
might expect that this profile does not give the best answer.

Among the results yielded by the Sevick model, one is particularly
surprising: at constant grafting density, the brush height is
predicted to decrease if the interface is curved more strongly.
Yet simple arguments indicate that on the contrary, the brush
height should increase. Suppose for a moment that the height does
not change when the curvature is changed. Curving the interface
more strongly then reduces the overall volume of the brush.
Two-body interactions are then more important, which tends to
swell the brush. The height will therefore stabilize at a somewhat
greater value.

A third approach considers the droplet curvature effect on the
emulsion stabilization by diblock copolymers \cite{Govorun}. They
study the small corrections due to curvature to the flat case in
the framework of the Alexander-de Gennes model, which is valid for
very weak curvatures.

To describe such concave brushes under large curvatures, one must
therefore keep in mind that the curved brush is some kind of a
compressed brush. Curving the interface indeed amounts to
increasing the brush concentration~: the corresponding compression
does not originate in an interaction at the brush edge, however,
but rather in some kind of lateral interactions (and the brush
profile is very different from that of a compressed planar brush,
as we indicate in Appendix~A). In this context, one result
obtained by Rabin and Alexander on compressed polymer brushes is
most important~\cite{RABIN}: they showed that, when the layer
thickness is forced to be smaller than that of a free brush (which
here corresponds to regimes~3 and~4), the main contribution to the
free energy of a compressed brush in good solvent is the osmotic
contribution. The elastic stretching energy of the chains is
comparatively smaller, even though they are still strongly
stretched (at least in regime~3). As a result, in the limit of
strong compressions (regimes~3 and~4, $R\ll aN\sigma ^{1/3}$), the
osmotic pressure is the mostly contributing term and is therefore
uniform over the whole brush volume. In other words, the monomer
concentration is uniform over the entire brush and the brush
structure is locally that of a semi-dilute solution of equal
overall concentration.

The size of the semi-dilute blobs is $\xi\simeq a\phi^{-3/4}$
(where $\phi=3aN\sigma /R$ is the volume fraction) and the free
energy of the whole brush is therefore~:
\begin{equation}
\label{F34} F_{3-4}^{\rm sph} \simeq \frac{R^3}{\xi^3}\simeq
(N\,\sigma)^{9/4}R^{3/4}
\end{equation}
This expression for the energy is valid well inside regime~3 or in
regime~4, {\it{i.e.}}, when the brush is strongly compressed
($R\ll L_{flat}$).

Our aim is to give a correct description of the chain structure
and free energy in regimes~2 and~3. The present work is organized
as follows.

We present in Section~2 a SCF approach for such concave brushes
valid for a near-theta solvent and an alternative convolution
method is outlined in Appendix~A. The purpose of Section 3 is to
give a scaling picture of concave brushes by refining SCF results
taking into account correlations between monomers. A discussion of
lateral fluctuations of grafted chains is given in Appendix~B,
where it is shown that they do not alter significantly the chain
free energy and thus do not endanger the SCF approach. In
Appendix~C, we give the results obtained in the cylindrical
geometry through a scaling approach.


\section{Self-consistent mean field description.}


In regimes~2 and~3, the polymer chains are strongly stretched and
it is thus possible to use the self-consistent field method
\cite{SEM85,MWC,ZHULINA}. We note $r$ the radial distance to the
spherical surface (defined by $r=0$).\\

The aim of this section is to determine the free energy $F$ by
solving for the monomer volume fraction $\phi (r)$ and for the
distribution $g(r)$ of chain-end monomers self-consistently. In
the classical limit of strong stretching~\cite{MWC}, the free
energy of the system is
\begin{eqnarray}
F &=& \int_{0}^{R}g(r_{0})\int_{0}^{r_{0}}\left[ \frac{3}{2a^2}
e(r,r_{0})+ v \frac{\phi (r)}{e(r,r_{0})}\right] drdr_{0}\nonumber
\\
&-&\frac{1}{2}\int_{0}^{R}\frac{4\pi(R-r)^{2}v}{a^{3}}\phi^{2}(r)dr
\label{F}
\end{eqnarray}
where $g(r_{0})dr_{0}$ is the number of chain ends in the
spherical shell of radius $R-r_{0}$ and width $dr_{0}$ and
$e(r,r_{0})=\left| \frac{dr}{dn}\right|$ is the local extension of
a chain whose free end is at $r_0$ (the variable $n$ denotes the
current monomer). The first term corresponds to a sum of
single-chain free energies in the potential $k_{B}T v \phi (r)$,
the second term corrects the fact that monomer-monomer
interactions have been double-counted. The normalization
conditions
\begin{equation}
\int_{0}^{r_{0}}\frac{dr}{e(r,r_{0})}=N
\label{norm1}
\end{equation}
\begin{equation}
\int_{0}^{R}4\pi (R-r)^{2}\phi (r)dr=4\pi R^2aN\sigma
\label{norm2}
\end{equation}
fix the number of monomers in one chain and in the whole system.
The mean field concentration is determined by $\delta
F/\delta\phi=0$, which yields
\begin{equation}
\phi (r)=\frac{a^{3}}{4\pi (R-r)^{2}}\int_{r}^{R}\frac{g(r_{0})}{%
e(r,r_{0})}dr_{0}
\label{SCF}
\end{equation}

We perform this calculation by two steps \cite{MWC}: first we
suppose that the brush height, $xR$, is fixed, and we minimize the
free energy given by Eq.~(\ref{F}) with respect to changes in the
field $\phi$. This yields a family of profiles $\phi_{min}$. Then
we determine this unknown parameter $x$ (with $0\leq x\leq 1$) by
minimization of $F[\phi_{min},x]$ which yields the brush height,
$\xs$, at equilibrium.\\
The ``equal-time" property can be applied for the self-consistent
field as was done previously in the case of a melt~\cite{SEM85} if
the following two assumptions hold~: {\em a)} there is no dead
zone containing no chain ends~\cite{DEADZONE}, which can be
written as $g(r_{0})\neq 0$ for $0<r_{0}/R<x$; and {\em b)} the
chain ends are not under tension~: $e(r_{0},r_{0})=0$. From the
equal-time argument, we know that the monomer concentration
profile is parabolic and we know the form of the local extension
of the chain~:
\begin{equation}
\phi_{min} (r)=\Phi\left[A-B\left(\frac{r}{R}\right)^{2}\right]
\Theta (r-xR) \label{parabolic}
\end{equation}
\begin{equation}
B=\frac{\pi^{2}R^{3}}{8 \sigma N^{3}a^{3}v}=\frac{10}{9}\left(
\frac{R}{R^*}\right)^{3} \label{B}
\end{equation}
where
\begin{equation}
R^*=\left(\frac{80}{9\pi^{2}}\right)^{1/3}a(v\sigma)^{1/3}N
\end{equation}
\begin{equation}
e(r,r_{0})= \frac{\pi}{2N} \sqrt{r_{0}^2-r^2} \label{stretch}
\end{equation}
where $\Theta$ is a step function, $\Phi=3aN\sigma /R$, $B$ is
obtained from the equal-time argument (Eq.~(\ref{norm1})) and $A$
is related to $x$ by (Eq.~(\ref{norm2}))
\begin{equation}
(3-3x+x^{2})xA=1+\frac{10}{9}\left( \frac{R}{R^*}\right)
^{3}x^{3}(1-\frac{3}{2}x+\frac{3}{5}x^{2}) \label{A}
\end{equation}
The minimization of the free energy per chain
$\widetilde{F}(x)=F[\phi_{min},x]/f$ with respect to changes in
$x$ under the three conditions {\em a)} $0\leq x\leq 1$, {\em b)
}$\phi (x^{-})\geq 0$ and {\em c)} $\phi (0)=\Phi A\leq 1$ is
illustrated in Figure~3. This gives rise to a distinction between
two cases depending on whether $R$ is smaller or larger than
$R^*$, where $R^*$ (see Eq.~(\ref{B})) is the critical radius for
which the sphere is entirely filled up by the brush:
\begin{itemize}
\item  for $R>R^*$ (regime~2), the minimum free energy $\widetilde{F}$
occurs for $\xs$ defined has the unique solution of the equation
\begin{equation}
x^{3}\left( \frac{4}{9}x^{2}-\frac{5}{3}x+\frac{20}{9}\right)
=\left( \frac{R^*}{R}\right)^{3}
\label{xstar}
\end{equation}
such as condition {\em a)} is fulfilled (Note that this condition
is equivalent to the smooth vanishing of the polymer concentration
at the edge of the brush \cite{Milner}). It remains a central
region of radius $R(1-\xs)$ in the center of the sphere which
contains pure solvent~;
\item  for $R\leq R^*$ (regime~3) the minimum of $\widetilde{F}$
occurs for $x=1$, there is no empty region and the concentration
at the center of the sphere is finite.
\end{itemize}
By varying $R$ from $\infty $ to $aN^{1/2},$ the limit of validity
of the SCF approach, the following scenario emerges.

\subsection{Regime~2 ($R>R^*$)}

In regime~2, the brush height $\xs$ is given by Eq.~(\ref{xstar})
and the monomer volume fraction (Eq.~(\ref{parabolic})) simplifies
into \cite{condition c} (Figure~4)
\begin{equation}
\phi (r)=\Phi B\,\left[x^{*2}-\left(\frac{r}{R}\right)^{2}\right]
\label{parabolic2}
\end{equation}
Equations (\ref{SCF}) and (\ref{parabolic2}) yield the chain end
distribution written in units of chains per unit length
$g(t)=\frac{g(r_0)a^2}{4\pi R\sigma}$ ~:
\begin{equation}
\label{gt}
\begin{tabular}{l}
$g(t)=Bx^{*2}t\left[3(2-3\xs +2x^{*2}) \sqrt{1-t^{2}}\,\,
-8x^{*2}(1-t^{2})^{3/2}\right. $\\ \qquad $\left.+6 \xs \arg \tanh
\sqrt{1-t^{2}}\,\, -9 \xs t^{2}\arg \tanh \sqrt{1-t^{2}}\right]$
\end{tabular}
\end{equation}
where $t=r_{0}/(\xs R)$. The distribution of chain ends can also
be expressed in terms of a number of chain ends per unit volume~:
$\rho(t)=\frac{g(t)}{3(1-t\xs)^{2}}$ (in units of chains per unit
volume, $\Phi /N$) for $0<t<1$. A typical probability density for
free chain ends is shown in Figure~5.

>From these, we readily derive the free energy per chain
$\widetilde{F}_{2}=\widetilde{F}(\xs)$ given by Eq.~(\ref{F})~:
\begin{equation}
\widetilde{F}_{2}=\frac{9}{10}\left(\frac{\pi^2}{4}\right)^{1/3}N
(v\sigma )^{2/3}\, \psi\left(\frac{R^*}{R}\right) \label{F2}
\end{equation}
where
\begin{eqnarray}
\psi(y) &=& \frac{10\, .\,4^{1/3}}{3(3-3\xs+x^{*2})}\left[
\frac{(9/10)^{1/3}}{4} \frac{y}{\xs}\right. \\ &+& \left.
\frac{(25/6)^{1/3}}{3}
\frac{x^{*2}}{y^{2}}(1-\frac32 \xs+ \frac35%
x^{*2})\right. \nonumber \\ &-& \left. \frac{(25/6)^{1/3}}{189}
\frac{x^{*5}}{y^{5}}(28-63\xs +\frac{201}{4}x^{*2}
-17x^{*3}+\frac{12}{5}x^{*4})\right]\nonumber \label{normF2}
\end{eqnarray}
In the entire regime~2, equation~(\ref{xstar}) shows that the
normalized brush height $\xs$ increases when the curvature $R^*/R$
increases, until it reaches $\xs=1$ for $R=R^*$ (Figure~6). In
other words, the brush height increases when the interface is
curved more strongly, as long as there is some free solvent in the
center. In the same way, it can be shown that the normalized free
energy $\psi$ (Eq.~(\ref{normF2})) increases monotonically at
constant grafting density when the curvature increases (Figure~7).

The osmotic pressure balances the elastic restoring force since
the polymer brush is at equilibrium under no external compression.
This readily explains~\cite{MWC} the fact that the monomer volume
fraction vanishes smoothly at the outer edge of the brush
(equation~(\ref{xstar}) is equivalent to $\phi(\xs)=0$).

Izzo and Marques~\cite{Marques} seem to have also used an SCF
approach for swollen concave brushes in this regime. They find the
same equation as~Eq.(\ref{xstar}) for the brush height, but their
free energy is different from the one given here. We do not
understand why at this stage.

In the limit of zero curvature, {\it i.e.} $R\rightarrow \infty$,
the brush height $L=\xs R$ given by Eq.~(\ref{xstar}) can be
expressed as~:
\begin{equation}
L_{flat}=\left(\frac{9}{20}\right)^{1/3}R^{*}=\left(
\frac{4}{\pi^{2}}\right)^{1/3}a(v\sigma)^{1/3}N
\end{equation}
which is the result of Milner, Witten and Cates \cite{MWC}.
Moreover, we have $\psi(0)=1$ and the energy Eq.~(\ref{F2})
becomes
\begin{equation}
\widetilde{F}_{2 flat}
=\frac{9}{10}\left(\frac{\pi^2}{4}\right)^{1/3}N (v\sigma )^{2/3}
\label{SCFflatbrush}
\end{equation}
which is also the classical result.

Going further, we can expand this result for very small
curvatures. Equation~(\ref{xstar}) yields the asymptotic value of
the brush height
\begin{equation}
L\simeq L_{flat}\left[
1+\frac14\,\frac{L_{flat}}{R}+\frac{29}{240}\,
\left(\frac{L_{flat}}{R}\right)^2 \right] \label{Rinfty}
\end{equation}
Note that a simple scaling method does not yield the correct value
for this asymptotic increase in the brush
height~\cite{RiAlexander}. We can also expand the free energy for
small curvatures
\begin{equation}
\widetilde{F}_{2}=\widetilde{F}_{2
flat}\left[1+\frac{5}{12}\frac{L_{flat}}{R}+\frac{73}{336}
\left(\frac{L_{flat}}{R}\right)^2\right]
\end{equation}
We observe that this is consistent with the results of Milner and
Witten~\cite{bending_moduli} who found for the convex geometry the
same development by replacing $R$ by $-R$. This is because, as
explained in~\cite{bending_moduli}, in the convex case, the dead
zone yields negligible terms  for small
curvatures~\cite{small_curv}.

In conclusion, for this region $2$, the SCF calculations show a
weak increase of the brush height when the interface is curved
more and more, which involves an increase of both elastic and
osmotic energy.

\subsection{Regime 3 ($aN^{1/2}\leq R\leq R^{*}$)\label%
{SCFcompresse}}

In regime~3, the brush height $L$ is fixed at $R$ ($x=1$). The
monomer volume fraction becomes \cite{condition c} (Figure~4)
\begin{equation}
\phi (r)=\phi (R)+\Phi
B\left[1-\left(\frac{r}{R}\right)^{2}\right] \label{parabolic3}
\end{equation}
where $\Phi(R)$ is the volume fraction at the center of the sphere
given by~:
\begin{equation}
\phi (R)=\Phi\left[1-\left( \frac{R}{R^{*}}\right)^{3}\right]
\label{sature}
\end{equation}
which increases when $R$ decreases. Thus, the monomer volume
fraction tends to become more uniform as the radius $R$ is
decreased.

The chain end distribution is ($t=r/R$)
\begin{equation}\label{gt2}
\begin{tabular}{l}
$g(t)=t\left[ -6\left(1-\frac{9}{10}B\right) \sqrt{ 1-t^{2}}\,\,
-8B(1-t^{2})^{3/2}\right. $ \\ \hspace{1cm}$
\left.+6\left(1+\frac{1}{10}B\right)\arg \tanh \sqrt{1-t^{2}}\,\,
-9Bt^{2} \arg \tanh \sqrt{1-t^{2}}\right]$
\end{tabular}
\end{equation}
As is shown in Figure~5, the major number of chain ends per unit
surface are located at the edge of the brush. The free energy per
chain is given by~:
\begin{equation}
\widetilde{F}_{3}=\left(\frac{9\pi^2}{10}\right)^{1/3}(v\sigma
)^{2/3} N \,\left[ \frac34 \frac{R^*}{R}+\frac16
\left(\frac{R}{R^*}\right)^2 -\frac{13}{756}
\left(\frac{R}{R^*}\right)^5 \right]
\label{F3}
\end{equation}
Mathematically, this free energy $\widetilde{F}_{3}$ is frozen at
$\widetilde{F}(x=1)$ because the equilibrium state $x=\xs >1$ is
out of reach, $x=1$ being a physical bound of the problem. This
situation is similar to the case of flat brush under compression
(a pressure is exerted at its edge): the concentration profile has
a truncated parabolic shape and its value at the edge of the
brush, Eq.~(\ref{sature}), increases when the compression is
stronger \cite{Martin}. As it has been shown by Milner {\it et
al.}, this means that there can be no ``interdigitation'' of
chains at $r=R$ \cite{Milner}.

At constant grafting density, $\widetilde{F}_{3}$ (Eq.~(\ref{F3}))
is a strongly increasing function of $R^*/R$ (Figure~7). A careful
analysis of this energy shows that for $R\ll R^{*}$, the last term
is negligible, it comes from the parabolic nature of the monomer
concentration. In the same way, from Eq.~(\ref{sature}) we can
reasonably consider that the monomer concentration is constant and
equal to $\Phi =3Na\sigma /R$. Therefore, the free energy can be
seen at the level of a simple Flory theory
\begin{equation}
\widetilde{F}_{3} \simeq
v\frac{aN^{2}\sigma}{R}+\frac{R^{2}}{Na^{2}}
\label{FcompresseSCF}
\end{equation}
where the first term is the osmotic pressure (at a mean-field
level) and the second one the elasticity of an ideal chain. For
$R\cong R^{*}$, these two energies balance, but when $R$ decreases
further, the excluded volume interactions strongly dominate and
the energy is practically equal to the osmotic pressure
(Figure~7). The point is that these two energies do not balance
because the brush height is geometrically constrained.


\section{Scaling refinements}


In the previous section, we used a self-consistent field approach
to study the conformation of a concave brush. Such an approach
describes both the average distribution of free chain ends and the
average position of any monomer, conditionally if the position of
the end-monomer of the chain to which it pertains is known.
It is known classically in two versions. The mean-field version
such as that exposed in the previous section provides a somewhat
erroneous expression for the brush free energy, which varies like
$\phi^2$. This flaw is intrinsic to the mean-field approximation
and is present also in the Flory approach.
A correct expression for the free energy can be obtained using
scaling concepts~\cite{PGGSCALING} and varies like $\phi^{9/4}$.
It was first used by de~Gennes in a scaling SCF approach for the
problem of polymer adsorption~\cite{PGGads}.  Milner, Witten and
Cates applied it in order to work out the conformation of a flat,
swollen polymer brush in good solvent~\cite{Milner}. The resulting
brush concentration profile obtained after minimization is
slightly different from that obtained through the mean-field SCF
calculation.
This is in contrast with the original Alexander-de~Gennes model
for the brush where both the mean-field and the scaling forms of
the free energy yield the very same brush conformation, whose
concentration profile is uniform.
In the SCF approach, the non-uniformity of the concentration is
the origin for the slight difference between mean-field and
scaling~\cite{MFSC}.
In this section, we present these scaling tools for the various
regimes of the spherical concave brush. For stretched chains
(regimes~2 and~3), we simply calculate the scaling free energy
from the brush conformation determined from the mean-field SCF
approach exposed in the previous section. Although it is not as
rigorous as the calculation of reference~\cite{Milner}, it should
be a sufficient approximation at least in regime~3 where the
concentration profile is essentially uniform, as we just
mentioned. In any case, the value of the free energy obtained in
this way in regimes~2 and~3 will be improved as compared to the
mean-field value obtained in the previous section.

We also describe more fully chain fluctuations around their
average conformation given by the SCF treatment. Confined chains
(regime~4) cannot be treated by the SCF theory~: we describe the
chain conformations and calculate the scaling free energy.

\subsection{Scaling refinements in regimes~2 and~3}

The scaling form of the free energy given in
reference~\cite{Milner} is~:
\begin{eqnarray}
\widetilde{F}^{sc} &\simeq& \frac12 \int_{0}^{R}
\frac{4\pi(R-r)^2}{a^3}\,\phi^{9/4}(r)dr \nonumber \\
&+&\frac{3}{2a^{2}}\int_{0}^{R}dr_0\,g(r_{0}) \int_{0}^{r_{0}}
e(r,r_0)\,\phi^{1/4}(r)dr  \label{Fsc}
\end{eqnarray}
where the factor $\phi^{9/4}$ in the osmotic term comes from the
$k_BT$ per blob Ansatz, while the $\phi^{1/4}$ factor in the
elastic term originates in the correction to the unperturbed
conformation of the chains which are swollen below the length
scale $\xi (r)=a\,\phi^{-3/4}$. The authors interpreted the
elastic term as a perturbation of each blob of size $\xi$~: it
corresponds to $kT\,(\Delta\xi/\xi)^2$ per blob, where $\Delta\xi$
is the average blob elongation.
But this term can be interpreted in a more intuitive way, using a
more precise description of the local chain conformation. As
mentioned briefly by the authors, at small length scales, chains
are like in a semi-dilute solution (swollen below $\xi$ and
Gaussian above $\xi$), while they are stretched at large length
scales.
In other words, the root mean square distance between two monomers
$n$ and $n'$ along the chain~(see Figure~8a) is given by~:
\begin{eqnarray}
\sqrt{\overline{(r-r')^2}}&\propto&(n-n')^{3/5}
\hs(\sqrt{\overline{(r-r')^2}}<\xi=a\phi^{-3/4})\nonumber\\
&\propto&(n-n')^{1/2}
\hs(\xi<\sqrt{\overline{(r-r')^2}}<\Lambda)\nonumber\\
&\propto&n-n' \hs(\sqrt{\overline{(r-r')^2}}>\Lambda)
\end{eqnarray}
where the intermediate length scale $\Lambda$, where this
transition occurs, is called the elastic blob (see
references~\cite{these,lambda})~:
\begin{equation}
\Lambda(r,r_0)\simeq a^2\left|\frac{dr}{dn}\right|^{-1}
\phi^{-1/4}
\end{equation}
At this length scale, the polymer chain is not an ideal random
walk any more~: instead, it has no choice but to move forward at
every step (exponent $1$). Hence, the chain is mainly a linear
string of these elastic blobs stretched along the radial
direction; the important point being that these blobs are defined
along one direction and that in the other two directions, blobs of
different chains overlap.

In the elastic term of equation~(\ref{F}), the degree of
stretching $e(r,r_0)$ must be replaced by an effective degree of
stretching $e_{\rm eff}(r,r_0)=a^2/\Lambda(r,r_0)$, which yields
the factor $\phi^{1/4}$ in equation~(\ref{Fsc}).
\newline
Note that the osmotic blob size $\xi(r)$ depends only on the
position $r$, whereas the size of an elastic blob $\Lambda(r,r_0)$
on a particular chain also depends on the local degree of
stretching of that chain, {\it{i.e.}}, on the position $r_0$ of
its free end.

In principle, the chain conformations should be obtained through
minimization of equation~(\ref{Fsc}), as in
reference~\cite{Milner}. Here, however, we simply use the volume
fraction $\phi(r)$ (equation~(\ref{parabolic2})
or~(\ref{parabolic3})), the free end distribution $g(t)$
(equation~(\ref{gt}) or~(\ref{gt2})), and the degree of stretching
derived through the mean-field method developed in the previous
section (Eq.~\ref{stretch}).

The free energy of the brush in regimes~2 and~3 can be computed
numerically from equation~(\ref{Fsc}) along the lines we have
indicated.
\newline
Here, for simplicity, we focus on the limit where $R\ll R^*$ (well
in regime~3). The free end distribution given by
equation~(\ref{gt2}) at zero order in $R/R^*$ then becomes~:
\begin{equation}
g(t)\simeq 6\,t\left\{ \arg\tanh\sqrt{1-t^2}-\sqrt{1-t^2}\right\}
\end{equation}
In the same way, at this order, $\phi(r)$ is constant where
$\phi(r)=\Phi=3aN\sigma/R$. With these two approximations, the
total free energy is equal to the first two terms of Eq.
(\ref{F3}) multiplied by $\phi^{1/4}$~:
\begin{eqnarray}
\widetilde{F} &\simeq& \frac32 \frac{aN^2v \sigma}{R}
\phi^{1/4}+\frac{3\pi^2}{80} \frac{R^2}{Na^2}\phi^{1/4}\\ &\simeq&
\left(\frac32 \right)^{1/4}
\left(\frac{9\pi^2}{10}\right)^{5/12}\sigma ^{5/6} N \,\left[
\frac34 \left(\frac{R^*}{R}\right)^{5/4}+\frac16
\left(\frac{R}{R^*}\right)^{7/4}\right] \nonumber
\end{eqnarray}
where the dominant contribution is osmotic (first term). The
second term is the elastic term (computed above for
$R/R^*\rightarrow 0$). The next nonzero contributions from the
osmotic and the elastic terms in the $R/R^*$ (or $B$) expansion
are smaller by a factor of order $(R/R^*)^3$.

Finally, in order to get simply the scaling dependence of
respectively the osmotic and elastic energy in regimes~2 and 3
(within unknown numerical coefficients of order unity), it is
possible to keep only the scaling dependence of blobs~: $\xi
\simeq a\Phi ^{-3/4}$ and $\Lambda \simeq N \Phi ^{-1/4}/R$ and to
assign the energy $k_{B}T$ per blob. Thus, in regime~2, we recover
the very simple estimate of the balance achieved by the stretched
chains first presented by Alexander \cite{Alexander} and de~Gennes
\cite{PGG}. The compressed brush in regime~3 is then analogous to
a flat brush in presence of a semi-dilute solution of monomer
volume fraction $\Phi_{b}=\Phi$ (see Figure 2). The parameter $R$
is related to a virtual external osmotic pressure $\Pi \sim \Phi
_{b}^{9/4}$ which compress the grafted layer \cite{these,PGG}.

\subsection{Confined chains (regime~4)}

The natural size of an $N$-monomer chain in a semi-dilute solution
of volume fraction $\phi$ is known to be $R(\phi)\simeq
aN^{1/2}\phi^{-1/8}=\xi(N/g)^{1/2}$ where
$\xi=ag^{3/5}=a\phi^{-3/4}$ is the osmotic blob size defined
earlier in this section. In regimes~2 and~3, when the sphere
radius is decreased, the chains remain stretched as long as their
end-to-end extension is larger than the elastic blob size
$\Lambda$. This breaks down when the radius of the sphere reaches
the natural size of the chain, {\it{i.e.}}, $R=R(\phi)$ or~:
\begin{equation}
R\cong a\sigma ^{-1/7}N^{3/7}
\end{equation}
For smaller sphere radii (which corresponds to regime~4, see
Figure~3), the grafted chains are confined. More precisely (see
Figure~8b), the chains are swollen at small length scales,
Gaussian at length scales between $\xi$ and $R$, and confined
(zero exponent) for larger numbers of monomers~\cite{GLOBULE}. The
transition between the Gaussian and the confined statistics
corresponds to $G=\phi^{1/4}R^2/a^2$ monomers. At scales larger
than $G$ monomers, because of the spherical boundary, the chain
has no choice but to stay in the same region of space instead of
performing a random walk~: globally, it is a string of $N/G$
overlapping blobs of size $R$. Using the same argument as earlier
in this section, we conclude that the compression contribution to
the overall energy can be estimated as~:
\begin{equation}
\widetilde{F}_{4el} \simeq \frac{N}{G} \simeq
N^{3/4}\sigma^{-1/4}(R/a)^{-7/4}
\end{equation}
This, however, is only a correction to the free energy, whose main
contribution is osmotic and is given by equation~(\ref{F34}).


\section{Concluding Remarks}


In this article, we have given both an SCF and a scaling study of
the structure and the associated free energy of polymer brushes in
concave geometries. A simple scaling description can be derived from
SCF calculations. The essential features are the following~: for
weak curvatures ($R>R^*\simeq 1.3\,L_{flat}$,
where $L_{flat}$ is the thickness of a planar brush
with the same grafting density), inwardly curved
brushes are a mere generalization of flat brushes~: the monomer
concentration decreases as we move off from the surface. A
specific scaling description which would account for the geometry
and for this slowly decreasing concentration
would yield nothing else
but constant blobs as in the Alexander-de~Gennes description of
flat brushes. This relatively poor scaling description can be
related to the fact that the chain end distribution spreads over
the entire layer. However, we argue that a self-similar profile of
inverted Daoud-Cotton type is not the correct scaling description.
Even if the variation of the thickness with $R$ does not allow to
bring to a firm conclusion, this approach does not lead to the
lowest free energy since not only it overestimates the actual free
energy but also it yields an increasing monomer concentration
which diverges in the center for $R\simeq L_{flat}$. At large
curvatures (keeping the grafting density constant), the monomer
concentration becomes progressively uniform, leading to the
structure of a squeezed semi-dilute solution where monomer-monomer
interactions predominate over chain stretching energy.
This description is consistent with the
fact that most chain ends are located at the center of the sphere,
just like in a compressed brush where they are located at the edge
of the brush. Finally, we also give elements to compute
numerically the free energy taking into account correlations
between monomers, by using the SCF results for the volume fraction
and the chain end distribution. We hope that the diagram of the
different regimes according to the values of $R$ and the grafting
density $\sigma$ will be checked experimentally in
microemulsions~\cite{Guenoun}.
\newline
As soon as $R\simeq L_{flat}$, the dissymmetry between convex and
concave cases, which originates
in the difference of available space for
the brush, becomes important and plays a non negligible role in
the stability of micelles and polymer decorated membranes. This
article brings the essential elements for the determination of
membrane spontaneous curvature, mean and Gaussian rigidity moduli
especially for decorated vesicles of high curvatures at
thermodynamic equilibrium (where the radius is of order of
$100\AA$~\cite{JOANNIC}). This theoretical work should also
illuminate the issue of the adsorption of polymers on rough
surfaces consisting of bumps and hollows of high
curvatures~\cite{LIGOURE}.
\newline
However, the question of membrane fluctuations with a short wavelength
$\lambda < L_{flat}$ still awaits consideration since
the range of steric
interactions between chains is then longer than $\lambda$.
The distribution and
conformation of grafted polymers should then be
altered only in the vicinity of the grafting points.
Indications of this
have been given experimentally by Singh {\it et al.}~:
the influence of surface roughness on polymer adsorption
is smaller when the roughness
has a very small wavelength~\cite{SINGH}.

\pagebreak

\begin{acknowledgement}

The authors wish to thank Alice Nicolas and Bertrand Fourcade for
useful discussions. We also thank E. Zhulina for pointing
out references~\cite{Zhulina1,Zhulina2} to us and for raising
the argument of the stabilizing concentration gradient at
the edge of the brush.

\end{acknowledgement}


\section*{Appendix~A: Convolution method
for a self-consistent field treatment of the concave brush}


We use here an alternate (and faster) method to calculate the free
energy in the SCF theory. As an example, we treat the regime 3.
\newline
We know from the works of Semenov and MWC that the potential
profile is parabolic, and that it is proportional to the volume
fraction:
\begin{equation}
\phi(r)=\Phi+\frac{1}{10}C -C\frac{r^2}{R^2}
\end{equation}
where $\Phi=3aN\sigma /R$ is the average volume fraction is
calculated by normalization and where $C=3\pi^2R^2/8a^2vN^2$ is
found from the ''equal-time argument". In regime~3, the volume
fraction has some finite value both  at the center
$\phi(R)\equiv\Phi-9/10C$. Taking advantage of these results, we
use the convolution method outlined in reference~\cite{CYPCONVO}
to determine the free energy (in mean-field). From this we
calculate the two-body excluded volume interaction free energy of
the brush~:
\begin{eqnarray}
F_{os} &=& \frac12\,\int 4\pi (R-r)^2dr\,
\frac{v}{a^3}\phi^2(r)\nonumber \\ &=& \frac{2\pi
R^3v}{a^3}\left\{ \frac13\Phi^2+\frac{13}{2100}C^2\right\}
\end{eqnarray}
Defining now $\phiflat(r)$ by
\begin{equation}\label{phiflat}
4\pi R^2\phiflat(r)\equiv 4\pi(R-r)^2\phi(r)
\end{equation}
as the density profile obtained by flattening out the concave
brush like a chopped pineapple slice, we compute the elastic
energy~\cite{CYPCONVO}~:
\begin{eqnarray}
F_{el} &=& 4\pi R^2 \frac{\pi^2}{8N^2a^5}\int_0^Rdr
\left\{-r^3\frac{d\phiflat(r)}{dr}\right\}\nonumber \\ &=&
\frac{\pi^3R^5}{2N^2a^5}\left\{
\frac{1}{10}\Phi-\frac{13}{700}C\right\}
\end{eqnarray}

Gathering all these results, we find the free energy in the mean
field approach (equation~(\ref{F3}))~:
\begin{equation}
F=\frac{2\pi}{3} \frac{R^3v}{a^3} \Phi^2 +\frac{\pi^3}{20}
\frac{R^5}{a^5N^2}\Phi -\frac{39\pi^5}{22400} \frac{R^7}{a^7vN^4}
\end{equation}
The free-end density profile is then obtained by a simple
convolution with the volume fraction profile~\cite{CYPCONVO}~:
\begin{eqnarray}
\frac{g(r_0)}{4\pi(R-r)^2}
&=&\int_{r_0}^R{\left\{-\frac{d\phiflat}{dr}(r)\right\}
\frac{r_0}{Na\sqrt{r^2-r_0^2}}dr}
\end{eqnarray}
where the last factor is the MWC free end distribution for a melt
brush (step-like volume fraction profile). Through integration, we
recover equation~(\ref{gt2}).


\section*{Appendix~B: Do lateral fluctuations ruin the SCF approach~?}


The SCF approach describes only a projection of the chain
conformation along the normal to the surface and ignores the chain
fluctuations parallel to the surface. For a brush grafted to a
flat interface, the lateral fluctuations can be described in very
simple terms since the chain performs a random walk in these
directions~: the square-averaged position of the $n$-th monomer
from the wall is $y\equiv an^{1/2}$ in the case of a melt, and
more generally $y\equiv an^{1/2}\phi^{-1/8}$ if $\phi$ is the
monomer volume fraction.

Such lateral fluctuations are not hindered in any way for a flat
brush since the potential is constant in the lateral directions.
In the case of a concave brush, however, when a chain fluctuates
laterally, it explores regions located closer to the grafting,
concave surface, and the potential is somewhat higher there. In
regions where this potential increase becomes of order $k_{B}T$,
we expect a reduction in the lateral fluctuations, which in turn
represents an additional reduction of entropy. In principle, the
potential should therefore be altered accordingly. We now estimate
this effect for the regime~3 (where it should be more important)
using a scaling description of the chain conformations.

Consider a point at a distance $r$ from the wall where the
potential has a certain value $V(r)$. Let us now consider the
potential as a function of the distance $z$ from the first point,
going perpendicularly to the radius~: from geometry, the new point
is closer to the wall by $\Delta r\simeq z^2/2(R-r)$. As a
consequence, the potential is increased by $\Delta r\,|dV/dr|$. By
considering correlations between monomers, the potential
equation~(\ref{B}) is modified according to \cite{Milner}:
$V(r')/k_{B}T=\phi^{5/4}(r')=\phi^{5/4}(r'(R))+3\pi^{2}/(8N^{2})(r'^
{2}(R)-r'^{2})$ where $dr'=\phi^{1/8}(r)dr$ and
$\phi(r')\equiv\phi(r)$. Hence, the variation of the potential in
the perpendicular direction is given by $(dV/dr')(dr'/dr)\Delta
r$.\\ In regime~3, the volume fraction is essentially uniform and
we have $r'\simeq r \phi^{1/8}$ thus
\begin{equation}
V(r,z)\simeq V(r)+k_{B}T\phi^{1/4} \frac{z^2\,r}{N^2\,(R-r)}
\end{equation}
In the direction of $z$, the chain evolves in the potential given
by the above equation. At small length scales, the potential
increase is weak and does not affect the chain conformation. At
large length scales, the chain is confined in this potential. At
the crossover between these two regimes, the typical distance $z$
that is actually explored by the chain is such that the monomers
required for a free random walk over that distance, each of which
being at a potential of order $V(r,z)$, correspond to an energy of
order $k_{B}T$ altogether. In other words, at a given radial
position $r$, the lateral fluctuations of the chain may not exceed
$z$ given by~:
\begin{equation}
k_{B}T\simeq
\frac{\overline{z}^2\phi^{1/4}}{a^2}[V(r,\overline{z})-V(r)]
\,\,\,{i.e.}\,\,\, \overline{z}\simeq
aN^{1/2}\phi^{-1/8}\left[\frac{R-r}{r}\right]^{1/4}
\end{equation}
As $\phi$ is essentially uniform, the lateral fluctuations depend
on position only through the last factor $[(R-r)/r]^{1/4}$. Near
the wall, the spontaneous fluctuations are given by~:
\begin{equation}
z\simeq an^{1/2}\phi^{-1/8}\hs(r\ll R)
\end{equation}
since this is still lower than $\overline{z}$. By contrast, close
to the center, the lateral fluctuations are given by~:
\begin{equation}
z\simeq aN^{1/2}\phi^{-1/8} [1-r/R]^{1/4}\hs(R-r\ll R)
\end{equation}
The energy involved in this effective chain confinement is
$k_{B}T$ per blob of size $z$, {\it{i.e.}}, per $G_z\simeq
(z/a)^2\phi^{1/4} \simeq N[(R-r)/R]^{1/2}$ monomers. Now, from the
SCF chain conformation $r=R\sin[\pi n/2N]$ (if the free end is at
the center $r=R$), we deduce that $(R-r)/R \sim (N-n)^2/N^2$.
Hence, the free energy due to the restriction of the lateral
fluctuations is given by~:
\begin{equation}
F_{\rm conf}\simeq \int_0^{O(N)}\frac{d(N-n)}{G_z} \simeq \ln N
\end{equation}
As a consequence, the free energy that corresponds to the
reduction of the lateral fluctuations is negligible. The
fluctuations should be included, however, to fully describe the
chain conformations near the center (for $R-r\ll R$).


\section*{Appendix~C: Cylindrical geometry.}


Following the same approach as in the Introduction, a scaling
diagram can be established for concave cylindrical brushes in
terms of the cylinder radius $R$ and the grafting density $\sigma$
(Figure~9). It should be noticed that two new regimes are present,
as compared to the spherical geometry~:
\begin{itemize}
\item{regime~1bis where the cylinder is so small ($R<aN^{3/5}$)
that mushrooms are confined by the cylinder and stretched along
the cylinder. This ensures the blob size equal to $R$ and their
length is $R_\parallel \simeq N/R^{2/3}$ \cite{PGGSCALING}.}
\item{regime~4 which is semi-dilute and where chains are not
laterally compressed. Their length is $R_\parallel \simeq
aN^{1/2}\phi^{-1/8}$.}
\end{itemize}
The limit between these two regimes is $R\simeq 1/N^3\sigma^3$.

\pagebreak




\newpage
\begin{flushleft}
{\large {\bf Figure Captions}}.
\end{flushleft}


\vspace{0.6cm}

{\bf Figure 1:} {\bf (a)} Conformation of a convex brush in the
Daoud-Cotton scaling description. The geometry ensures the
equilibrium between elastic and osmotic forces. The same picture
transposed to concave surfaces {\bf (b)} is used in
references~\cite{Lipowsky,SEVICK,Zhulina1}. The inset shows the
monomer volume fraction profile.

\vspace{0.6cm}

{\bf Figure 2:} Various regimes for a spherical concave grafted
polymer brush. The inverse of curvature is $R$ and the grafting
density $\sigma$. A schematic representation of the brush is shown
using blobs defined in Section 3. This figure is only qualitative.
(1: mushrooms, 2: weak concave brush, 3: compressed brush, 4:
collapsed brush, in gray: unphysical regions)

\vspace{0.6cm}

{\bf Figure 3:} Variations of the free energy per chain
$\widetilde{F}(x)$ for three different values of $R/R^{*}$ for
$N=1000$, $\sigma =0.03$ ($R^{*}/a=300.07)$: from bottom to top
$R=2R^{*}$ (region $2$); $R=R^{*}$; $R=0.8R^{*}$ (region $3$). The
dashed part of the curve corresponds to an unphysical
situation~\cite{condition b}.

\vspace{0.6cm}

{\bf Figure 4:} Concentration profile of the brush. {\bf(a)} Large
sphere ($R=2~R^*$, regime~2) : the profile is parabolic and the
concentration vanishes at some finite distance $\xs$ from the
surface of the sphere (see Eq.~(\ref{parabolic2})). {\bf(b)} At
$R=R^*$, the concentration vanishes precisely at the center of the
sphere. {\bf(c)} Small sphere ($R=0.8~R^*$, regime~3) : truncated
parabolic profile with a finite concentration at the center of the
sphere (see Eq.~(\ref{parabolic3})).

\vspace{0.6cm}

{\bf Figure 5:} Variation of the normalized chain end density
distribution $\rho$ vs. $r_{0}/R$ for two different radii,
$R=1.27\,R^*$ (regime~2) and $R=0.75\,R^{*}$ (regime~3) ($N=1000$,
$f=600$).

\vspace{0.6cm}

{\bf Figure 6:} Variation of the solution $\xs$ of equation (15)
vs the normalized curvature $R^*/R$. The inset shows the variation
of the normalized brush height $L/L_{flat}$ vs the normalized
curvature of the sphere $R^*/R$.

\vspace{0.6cm}

{\bf Figure 7:} Variation of the scaled total free energy per
chain $\psi$ (see Eq.~(19)) in regimes~$2$ and $3$ vs the
normalized curvature $R^*/R$. Are also represented the elastic
($\psi_{el}$) and osmotic ($\psi_{os}$) parts (where
$\psi=\psi_{os}+\psi_{el}$)

\vspace{0.6cm}

{\bf Figure 8:} Mean square distance between two monomers from the
same chain (radial component), as a function of the number of
monomers between them (logarithmic scales). {\bf{(a)}} In
regime~3, for a given average degree of stretching $|dr/dn|$ and
volume fraction $\phi$ (thin line), the thick line is the scaling
picture. At long distances, the dependence is linear, which
reflects the fact that the chain is stretched (exponent $1$). At
smaller distances, the chain takes the conformation it would have
in a semi-dilute solution of volume fraction $\phi$ : it is
swollen at small length scales (exponent $3/5$) and it is Gaussian
at length scales larger than $\xi=a\phi^{-3/4}$ (exponent $1/2$).
{\bf{(b)}} In regime~4, the chains are compressed~: the SCF
approach does not apply and the scaling approach yields the
following picture. For a given volume fraction $\phi$, all chains
behave alike. They are swollen at small length scales (exponent
$3/5$) and Gaussian at length scales larger than
$\xi=a\phi^{-3/4}$ (exponent $1/2$ like in a semi-dilute
solution). A $G$-monomer strand spans over the entire interior of
the sphere of size $R$ and thus larger strands do not extend any
further~: they are confined (exponent $0$).

\vspace{0.6cm}

{\bf Figure 9:} Various regimes for a cylindrical concave grafted
polymer brush. The inverse of curvature is $R$ and the grafting
density $\sigma$. This figure is only qualitative. (1: mushrooms,
1bis: compressed mushrooms, 2: weak concave brush, 3: compressed
brush, 4: semi-dilute brush)



\begin{thebibliography}{10}
\bibitem{DEADZONE}
The fact that there is no dead zone for the extremities is crucial
for applying this SCF method to concave brushes. A dead zone is
present for a convex geometry~\cite{SEM85}.
\bibitem{condition c}  In principle, the condition {\em c)}
is not fulfilled for every value of parameters ($\sigma ,R)$. In
region $2$, the condition {\em c)} is rewritten $L\leq
(2\sqrt{2}/\pi \sqrt{3})v^{1/2}N$ which is valid for every radii
$R$ if $\sigma \leq \sigma _{1}=(\sqrt{6}/5\pi )v^{1/2}.$ In
region $3$, if $\sigma \leq \sigma _{1}$, the condition {\em c)}
is fulfilled for a range of radii $R_{1}\leq R\leq R^{*}$ where
$R_{1}=(4\sqrt{5}/3\pi )v^{1/2}N [ \cos( \frac{1}{3}\arctan
\sqrt{(\frac{\sigma _{2}}{\sigma} )^{2}-1}) -\sqrt{3}\sin(
\frac{1}{3}\arctan \sqrt{(\frac{\sigma _{2}}{\sigma})^{2}-1})] $
and $\sigma _{2}=(8\sqrt{5}/27\pi )v^{1/2}$. For $\sigma
>\sigma _{1}$, the parabolic solution for the monomer volume
fraction is not valid and it has to be set equal to $1$ near the
surface. However this remains a marginal effect and concerns a
very small range of values of the parameters ($\sigma ,R)$, since
the mean field treatment is valid in the limit of weak $v$ and at
moderately high $\sigma $ (formally for $v^{2}\ll \sigma \ll 1)$.
\bibitem{RiAlexander}
At a scaling level of description, the brush height and the free
energy are constant for the whole regime~$2$ within numerical
coefficients of order unity. An attempt at calculating the slight
increase of the brush height with decreasing sphere radius can be
made by minimizing the Alexander free energy: $$
\widetilde{F}\cong \Phi^{1/4}\frac{L^{2}}{a^{2}N}+N\Phi^{5/4} $$
where $\Phi$ is related to the brush height $L$ through $\Phi=4\pi
R^{2}aN\sigma /V$, where $V=4/3\pi[R^3-(R-L)^3]$ is the volume
occupied by the brush. At first order, the brush height increment
can be written as $L/L_{flat}-1\cong \frac13 aN\sigma^{1/3}/R$.
This result is incorrect, as can be seen by comparison with
equation~(\ref{Rinfty}). Indeed, the monomer concentration is
assumed to be uniform in this scaling description. Hence, the
concentration near the edge of the brush, which contributes
slightly more to the present effect of curvature, is
overestimated. That is the reason for the overestimated factor
$\frac13$ instead of $\frac14$. The correction to the scaling law
$\Phi\cong \sigma^{2/3}$ is, at first order $$ \Phi \cong
\sigma^{2/3}+\frac{2}{3}\left(\frac{aN\sigma}{R}\right) $$ Note
that this development is valid in the regime 2 that is for $R\gg
aN\sigma^{1/3} \cong R^{*}$.
\bibitem{small_curv} We recover the old result of stability of
symmetrically polymer decorated membrane under fluctuations of
large wavelength (at a mean field level)~\cite{bending_moduli}.
However, for surfactant vesicles, the grafting is at thermodynamic
equilibrium. Thus, in general, the grafting density is not
symmetric~\cite{JOANNIC}. This explains why polymer decorated
membranes form spontaneously vesicles.
\bibitem{MFSC} Since the exponents for the volume fraction $\phi$
in the elastic and in the osmotic terms of the free energy are
slightly higher in scaling than in mean-field, the brush tends to
avoid presenting any high concentration region~: it is rather
distributed more evenly. For example, in the parabolic profile
presented in reference~\cite{Milner}, the scaling profile is
slightly more flattened in the high $\phi$ region near the wall
than the mean-field profile, and as a compensation, it is slightly
more concentrated in the low volume fraction region near the edge
of the brush.
\bibitem{GLOBULE}
This is somewhat similar to the collapsed structure
of a chain in a poor solvent,
which is Gaussian both below and above $\xi$,
and confined for number of monomers larger than $G$.
In the case of the globule in poor solvent,
the expressions for $\xi\simeq a\phi^{-1}$
and $G\simeq (R/a)^2$ differ somewhat
from the situation studied here.
\end{thebibliography}
\end{document}